\documentclass[a4paper,11pt]{article}
\pdfoutput=1 

\usepackage{jheppub} 

\usepackage{enumerate}
\newcommand{\varA}[1]{{\operatorname{#1}}}

\title{Improved axion haloscope search analysis
}
\author[a,b]{S. Ahn,}
\author[b]{S. Lee,}
\author[b]{J. Choi,}
\author[b,1]{B. R. Ko,\note{Corresponding author.}}
\author[a,b]{and Y. K. Semertzidis\,}
\affiliation[a]{Department of Physics, Korea Advanced Institute of
  Science and Technology (KAIST), Daejeon 34141, Republic of Korea}
\affiliation[b]{Center for Axion and Precision Physics Research
  (CAPP), Institute for Basic Science (IBS), Daejeon 34051, Republic
  of Korea}
\emailAdd{brko@ibs.re.kr}

\abstract{
  One of the most significant and practical figures of merit in axion
  haloscope searches is the scanning rate, because of the unknown
  axion mass. Under the best experimental parameters, the only way to
  improve the figure of merit is to increase the experimentally
  designed signal to noise ratio in the axion haloscope search
  analysis procedure. In this paper, we report an improved axion
  haloscope search analysis using the data taken by the CAPP-8TB
  haloscope. By correcting for the background biased by the background
  parametrizations in the presence of axion signals, we realized a
  signal to noise ratio efficiency of about 100\%. Given the axion
  haloscope search analyses to date, the scanning rate can be improved
  by 21\%, with about a 10\% improvement in the signal to noise
  ratio. This improvement is another low cost innovation in axion
  haloscope searches, where all the experimental parameters are
  currently at their best.
}


\begin{document}

\maketitle
\flushbottom

\section{Introduction}
The axion~\cite{AXION} is an elementary particle considered to result
from a breakdown in a new symmetry first proposed by Peccei and Quinn
(PQ symmetry)~\cite{PQ} to solve the strong $CP$ problem in the
Standard Model of particle physics ~\cite{strongCP}. This particle is
massive, stable, and is born cold by the PQ symmetry breaking.
This makes the axion one of the most promising candidates for cold
dark matter, which constitutes about 85\% of the matter in the
Universe according to cosmological measurements and the standard model
of Big Bang cosmology~\cite{PLANCK}.

The method of searching for the axion proposed by
Sikivie~\cite{sikivie}, also known as the axion haloscope search,
involves a microwave resonant cavity with a strong static magnetic
field that induces axions to convert into microwave photons.
Using the resonant cavity, the axion signal power can be
enhanced when the axion mass $m_a$ matches the resonant frequency
of the cavity mode $\nu$, $m_a=h\nu/c^2$. Because the axion mass is
unknown, however, the resonant cavity has to be tunable, to allow axion
haloscope searches to scan all frequencies corresponding to possible
axion masses.
Because of this frequency scanning procedure, the most significant and
practical figure of merit in axion haloscope searches is the scanning
rate~\cite{scanrate}
\begin{equation}
  \frac{d\nu}{dt}=
  \frac{\nu b_a}{Q_L}
  \Big(\frac{1}{\rm SNR_{target}}\Big)^2
  \Big(\frac{P^{a\gamma\gamma}_a}{P_n}\Big)^2
  \propto\frac{g^4_{a\gamma\gamma}B^4V^2C^2Q_L}{{\rm SNR^2_{target}} T^2_n}
  \label{scanrate}
\end{equation}
for a target signal to noise ratio ${\rm SNR_{target}}$.
$P^{a\gamma\gamma}_a$ is the axion signal power proportional to
$g^2_{a\gamma\gamma}B^2 VC Q_L$~\cite{sikivie, scanrate}, where
$g_{a\gamma\gamma}$ is the axion-photon coupling strength, $B$ is a
static magnetic field provided by magnets in the axion haloscopes, $V$
is the cavity volume, $C$ is a form factor representing the overlap
between the electric field of the cavity mode and the static magnetic
field whose general definition can be found in Ref.~\cite{EMFF_BRKO},
and $Q_L$ is the loaded quality factor of the cavity mode. $P_n$ is
the noise power proportional to the noise temperature $T_n$ and the
axion signal window $b_a$.

From the radiometer equation~\cite{DICKE}, the SNR is
\begin{equation}
  {\rm SNR_{designed}}=\frac{P^{a\gamma\gamma}_a}{\sigma_{P_n}},
  \label{snr}
\end{equation}
where $\sigma_{P_n}$ is the noise power fluctuation.
The subscript ``designed'' stands for experimentally designed,
thus the ${\rm SNR_{designed}}$ must be designed to be
the same as the $\rm SNR_{target}$ or the $\rm SNR_{target}$ must be
set to be the ${\rm SNR_{designed}}$ in order to have an axion
haloscope search in an experimentally designed time.
However, generally the experimentally achieved signal to noise ratio
${\rm SNR_{achieved}}$ is smaller than the designed one by
holding the relation
${\rm SNR_{achieved}}=\epsilon_{\rm SNR}{\rm SNR_{designed}}$,
where $\epsilon_{\rm SNR}$ is the reconstruction efficiency of the
${\rm SNR_{designed}}$ in an axion haloscope search analysis procedure
whose values vary from about 50 to 90\% depending on the analysis
strategy~\cite{ADMX, HAYSTAC, simple_ACTION}.

Having said that, the scanning rate guides the following two cases.
\begin{enumerate}[(I)]
\item For axion haloscope searches to achieve the target
  sensitivity or ${\rm SNR_{target}}$ in an experimentally designed
  time, the experimental parameters, $B$, $V$, $C$, $Q_L$, and $T_n$
  have to be designed to meet the condition ${\rm SNR_{achieved}}={\rm SNR_{target}}$,
\item For axion haloscope searches to achieve the target
  sensitivity or ${\rm SNR_{target}}$, axion haloscopes have to take
  data until the condition ${\rm SNR_{achieved}}={\rm SNR_{target}}$
  is satisfied, even under the experimental parameters, $B$, $V$, $C$,
  $Q_L$, and $T_n$ at their best.  
\end{enumerate}
In both cases, the figure of merit of the experiments can be enhanced
by improving the ${\rm SNR_{achieved}}$ or $\epsilon_{\rm SNR}$, which
results in more sensitive results for (I) and shorter data acquisition
periods for (II).

In this paper, we report an improved axion haloscope search analysis
using the data taken by the CAPP-8TB
haloscope~\cite{CAPP-8TB-PRL}. Using the improved axion haloscope
search analysis in this paper, the figure of merit of axion haloscope
searches can be effectively increased by about 21\%.

\section{Axion haloscope search analysis strategy}
The simplest axion haloscope search analysis strategy is the one-bin
search that was employed in Ref.~\cite{simple_ACTION}, where all the
axion signal power belongs to a single frequency bin width
corresponding to the axion signal window $b_a$, if axions are there.
The price for the simplicity of this one-bin search is a very low
reconstruction efficiency of the axion signal power
$\epsilon_{P^{a\gamma\gamma}_a}$.
As described in Refs.~\cite{ADMX, simple_ACTION},
we lose about 20\% of the signal power by choosing a signal window to
get an optimized SNR, and an additional 20\% from the frequency
binning choice.

\begin{figure}[h]
  \centering
  \includegraphics[width=0.7\textwidth]{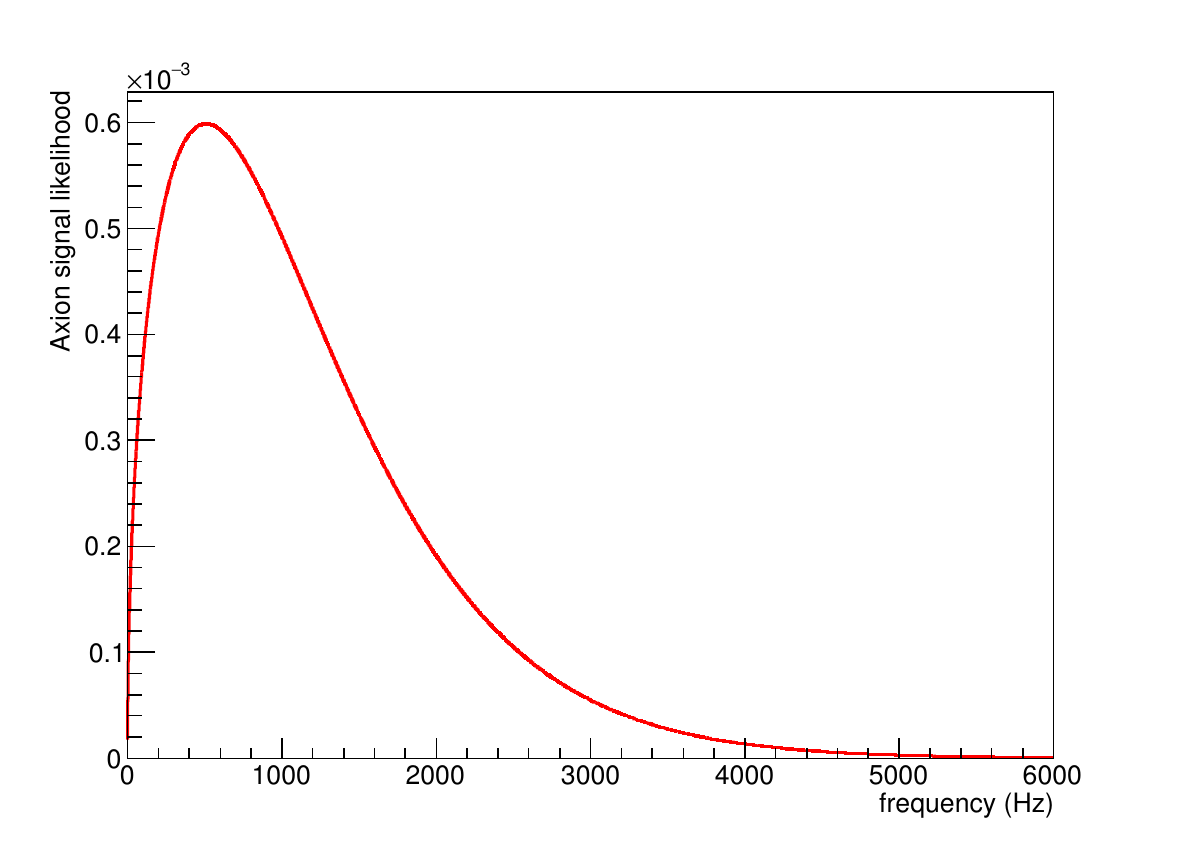}
  \caption{Axion signal likelihood distribution for an axion mass of
    1.625 GHz, where the axion mass is downconverted to zero. We can
    retain about 99.9\% of the total signal power by taking the signal
    window of 5000 Hz or, equivalently, 10 frequency bins with a
    resolution bandwidth of 500 Hz.}  
  \label{FIG:AXION_SHAPE}
\end{figure}

The other strategy is the multi-bin co-adding
search developed by the Axion Dark Matter eXperiment
(ADMX)~\cite{ADMX}, where all the axion signal power is distributed over
the multi frequency bins obeying an axion signal shape~\cite{turner}
(also shown in Fig.~\ref{FIG:AXION_SHAPE}). The multi frequency bin
width also corresponds to the axion signal window $b_a$. The multi-bin
co-adding search overcomes the inefficiency caused by the selection of
frequency binning.

The multi-bin co-adding search with signal likelihood weighting
developed by the Haloscope At Yale Sensitive To Axion Cold dark matter
(HAYSTAC)~\cite{HAYSTAC} overcomes the inefficiency resulting from the
choice of signal window.

As reported in~\cite{ADMX, HAYSTAC, simple_ACTION}, however, the
background subtraction with a 5-parameter
fit~\cite{ADMX, simple_ACTION} or a Savitzky-Golay
filter~\cite{HAYSTAC} also generates an additional inefficiency of about
20\% in the SNR reconstruction in the axion haloscope search analysis
procedure. The HAYSTAC analysis procedure recovered this inefficiency
using a single frequency-independent numerical factor, resulting in an
improvement of about 8\% in $\epsilon_\varA{SNR}$~\cite{HAYSTAC}.
Another low cost innovation in axion haloscope searches might be to
improve the inefficiency induced by the background subtraction, and
that is the main contribution of this paper.

\section{Axion haloscope search analysis procedure}\label{SEC:PROCEDURE}
The axion haloscope search analysis procedure can be divided into the following
three steps:
\begin{enumerate}[]
\item Step-1 ; background parametrization for the background
  subtraction\footnote{background parametrization also for the
    filtering of the individual power spectrum~\cite{HAYSTAC}},
\item Step-2 ; combining all the power spectra as a single power
  spectrum, taking into account the overlaps among the power spectra,
\item Step-3 ; constructing a ``grand power spectrum'' by co-adding
  multi-bins with signal likelihood weighting.
\end{enumerate}
After the background subtraction in Step-1, the normalized power
excess at each stage is likely to follow a Gaussian distribution in
the absence of axion signals, because the background in axion
haloscope searches is very like stationary Poisson noise. The mean of
the Gaussian depends on the expected SNR, while its width should be
unity if the power excess errors are estimated correctly. All of the
effort in this work is focused on obtaining such a Gaussian
distribution in each step, from Step-1 to Step-3.

\section{Data and parameters}
Here we describe the data used for this work. It consists of:
\begin{enumerate}[(i)]
\item experimental data from the CAPP-8TB experiment~\cite{CAPP-8TB-PRL},
\item 5000 simulated axion haloscope search experiments with a flat
  background and axion signals at a particular frequency, referred to
  as ``flat background simulation data'',  
\item 5000 simulated axion haloscope search experiments with the
  CAPP-8TB backgrounds and axion signals at a particular frequency,
  referred to as ``CAPP-8TB simulation data'',  
\item 5000 simulated axion haloscope search experiments with the
  CAPP-8TB backgrounds only, referred to as ``background only
  simulation data'',
\item other 4 sets of 5000 simulated axion haloscope search
  experiments with the CAPP-8TB backgrounds, where each set has axion
  signals at different frequencies, thus has different SNRs, referred
  to as ``other CAPP-8TB simulation data''.
\end{enumerate}

Here we also employ some assumptions and parameters adopted by the
CAPP-8TB axion haloscope search~\cite{CAPP-8TB-PRL}. We assume the
axions have an isothermal distribution, thus distribute over a boosted
Maxwellian shape, as shown in Fig.~\ref{FIG:AXION_SHAPE}, with the
following parameters: an axion rms velocity of about 270 km/s and the
Earth rms velocity of 230 km/s with respect to the galaxy
frame~\cite{turner}. We took 5000 Hz as the signal window of the
axions considered in CAPP-8TB. We can then retain about 99.9\% of the
total signal power as shown in Fig.~\ref{FIG:AXION_SHAPE}. After
Step-1, the five nonoverlapping frequency bins in each
background-subtracted power spectrum were merged so that the
resolution bandwidth (RBW) became 500 Hz, which we refer to as
``Step-1.5''. An RBW of 500 Hz was chosen not only to retain the axion
signal shape shown in Fig.~\ref{FIG:AXION_SHAPE}, but also to avoid
unnecessarily inordinate calculation time. The Step-2 and Step-3
procedures then went through with the RBW of 500
Hz~\cite{CAPP-8TB-PRL}. With our 10 co-adding bins in Step-3,
therefore, $P^{a\gamma\gamma}_a$ is almost 100\% retained.

\section{Validations}
\begin{figure}[h]
  \centering
  \includegraphics[width=0.7\textwidth]{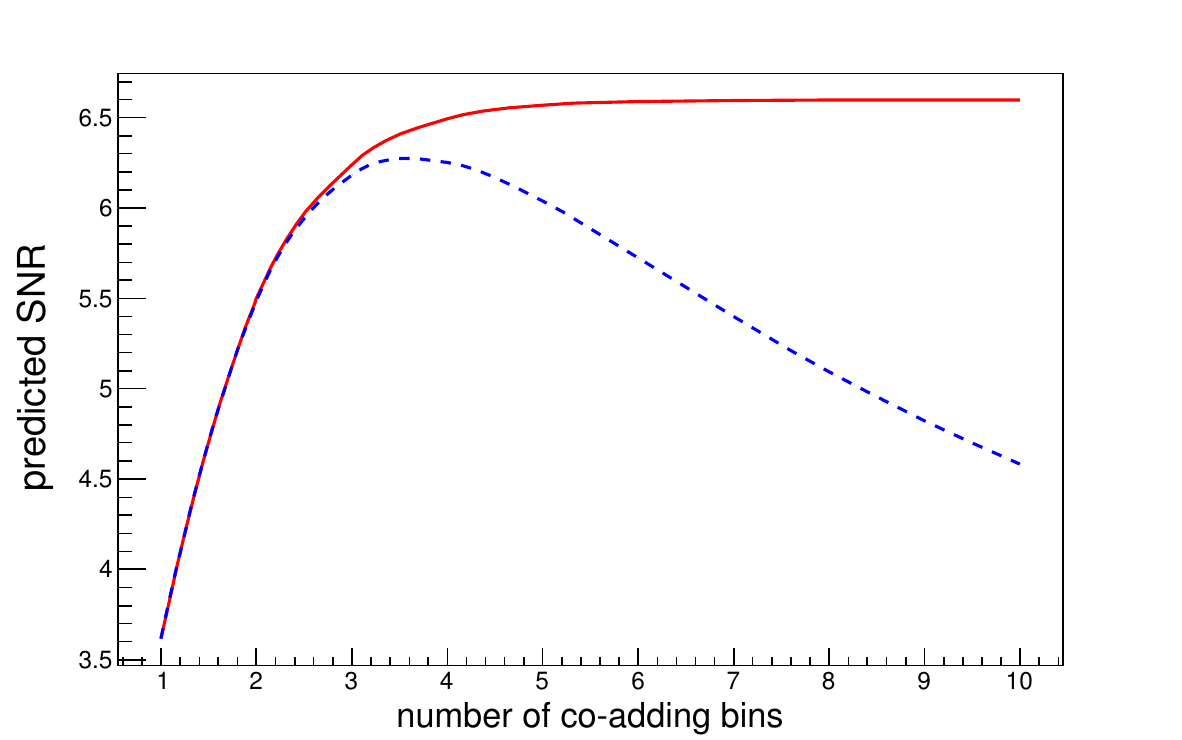}
  \caption{Predicted SNR as a function of the number of co-adding
    bins. Solid (red) and dashed (blue) lines show SNRs with and
    without the signal weighting applied, respectively.}  
  \label{FIG:SNR_VS_NBINS}
\end{figure}
Using the flat background simulation data, first, we validated our
understanding of the axion haloscope search analysis procedure,
especially the signal weighting in the Step-3 procedure as well as our
simulation data, particularly the signal injections that must reflect our
cavity responses. With a flat background, one can easily predict solid
SNRs with little background dependence in the Step-2
procedure. Figure~\ref{FIG:SNR_VS_NBINS} shows the predicted SNR as a
function of the number of co-adding bins with and without the signal
weighting. The weighting factors were obtained by integrating
over the relevant regions of the axion signal likelihood shown in
Fig.~\ref{FIG:AXION_SHAPE}.

The flat background simulation data was fed through the procedure
described in Sec.~\ref{SEC:PROCEDURE}. The backgrounds were subtracted
without any fit parametrizations. All the background-subtracted power
spectra were then combined as a single power spectrum. In constructing
a grand power spectrum, each power spectral line in the single power
spectrum was weighted accordingly~\cite{HAYSTAC} using the axion signal
likelihood shown in Fig.~\ref{FIG:AXION_SHAPE}.
\begin{figure}[h]
  \centering
  \includegraphics[width=0.7\textwidth]{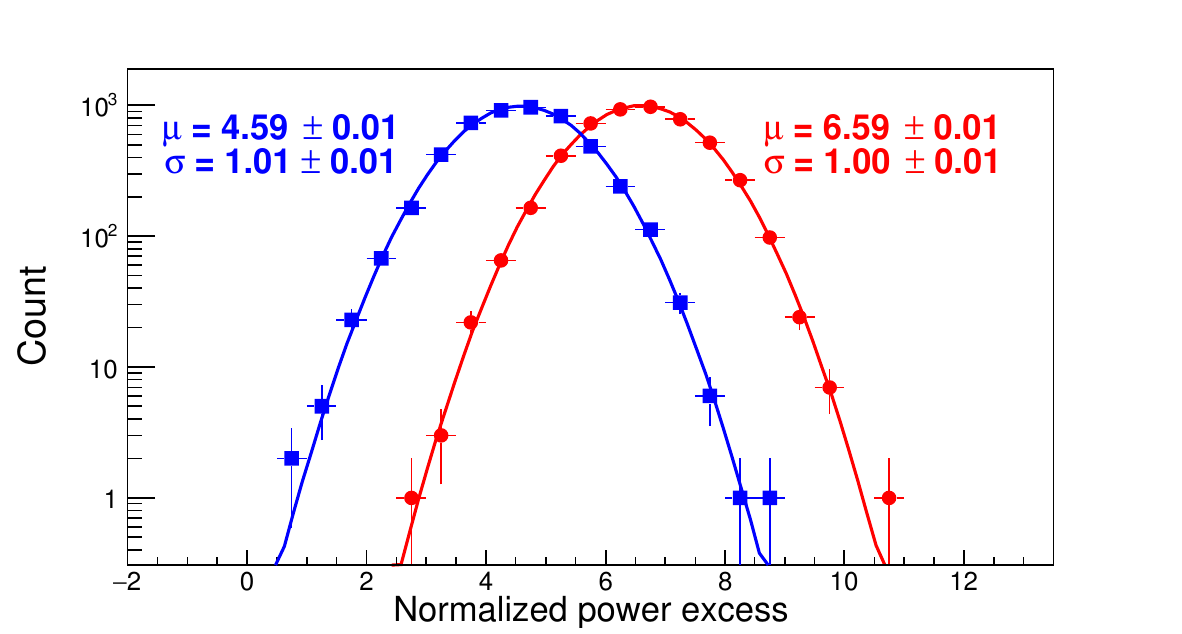}
  \caption{Distributions of the normalized power excess from a
    particular frequency bin where we put simulated axion signals,
    with (circles) and without (rectangles) the signal weighting
    applied. Lines are a Gaussian fit.}  
  \label{FIG:CAPPSIM_FLAT}
\end{figure}
It was further normalized by the corresponding noise fluctuation,
which was also weighted according to the axion signal
shape~\cite{HAYSTAC}.
With a signal window of 5000 Hz or,
equivalently, 10 co-adding bins, Fig.~\ref{FIG:CAPPSIM_FLAT} shows two
distributions of the normalized power excess from a particular
frequency where we put simulated axion signals, with (circles) and
without (rectangles) the signal weighting,
respectively. Figure~\ref{FIG:CAPPSIM_FLAT} also shows the Gaussian
fit results, including the means ($\mu$) and widths ($\sigma$) of the
distributions. Both means follow the predicted SNRs shown in
Fig.~\ref{FIG:SNR_VS_NBINS} and the widths are unity as they should
be.

Having demonstrated our solid understanding of the simulation data and
axion haloscope search analysis procedure from the flat background
simulation data, we applied the same analysis procedure to the
CAPP-8TB simulation data as well as the CAPP-8TB experimental data,
but with a 5-parameter fit~\cite{ADMX} for the background
parametrization and subtraction.
The same procedure except for the background subtraction using the
simulation input background functions, i.e., a perfect fit, was also
applied just to the simulation data to determine the inefficiency
resulting from the background
subtraction~\cite{ADMX, HAYSTAC, simple_ACTION}.
\begin{figure}[b]
  \centering
  \includegraphics[width=1.0\textwidth]{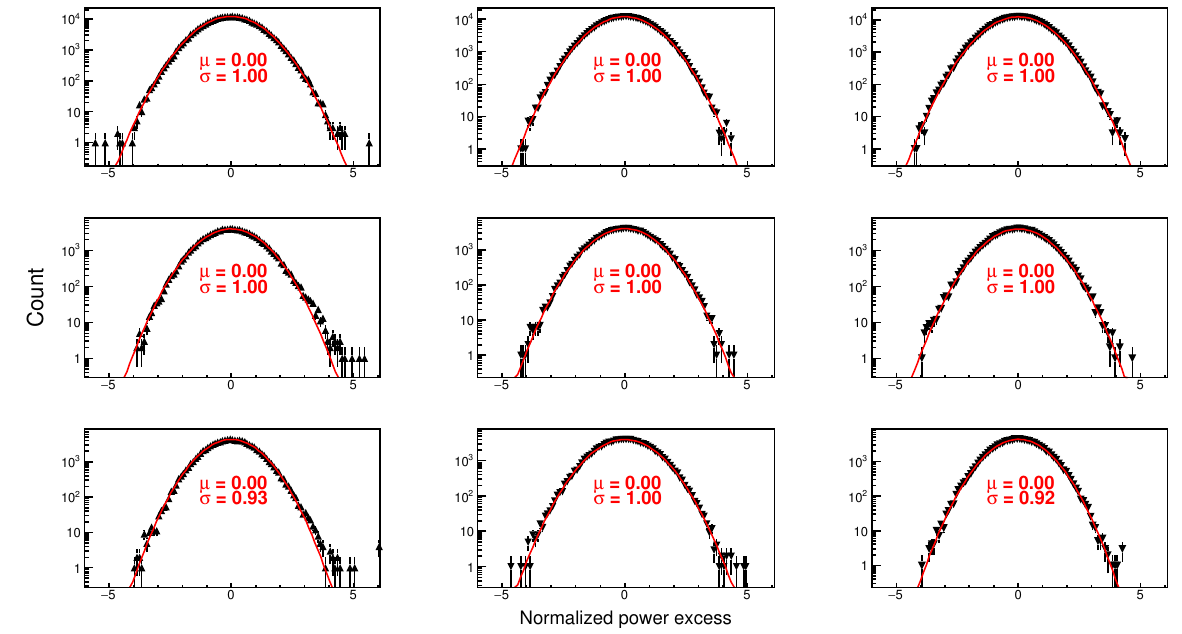}
  \caption{Distributions of the normalized power excess from the
    CAPP-8TB experiment (triangles in the 1st column), and from a
    single simulated axion haloscope search experiment with CAPP-8TB
    backgrounds and axion  signals at a particular frequency (inverted
    triangles in the 2nd and 3rd columns). Distributions in the 2nd
    column were obtained after subtracting the background with perfect
    fit, while those in the 1st and 3rd columns were with a
    5-parameter fit. From top to bottom row, they are the
    distributions after the Step-1, Step-2, and Step-3 procedures,
    respectively. Lines are a Gaussian fit resulting in $\mu$ and
    $\sigma$ with negligible fit errors of
    $\mathcal{O}(10^{-3})$~\cite{SMALLERROR}.}  
  \label{FIG:CAPPSIM_SINGLE}
\end{figure}
The normalized power excess distributions from the CAPP-8TB
experimental data are shown as triangles in the 1st column of
Fig.~\ref{FIG:CAPPSIM_SINGLE}. The distributions from a single
simulated axion haloscope search experiment with CAPP-8TB backgrounds
and axion signals at a particular frequency are shown in the 2nd and
3rd columns of Fig.~\ref{FIG:CAPPSIM_SINGLE} as inverted triangles.
The distributions in the 1st and 3rd columns were obtained after
subtracting the background with a 5-parameter fit, while in
the 2nd they were obtained with a perfect fit.
From top to bottom, they are the distributions after the Step-1,
Step-2, and Step-3 procedures, respectively.
The distributions after the Step-3 procedure are narrower than unity
for both the experimental and simulation data with a 5-parameter
background parametrization. This has been previously observed in axion
haloscope search experiments~\cite{ADMX, HAYSTAC, simple_ACTION}.
Without shrinking the distribution after the Step-3 procedure with
perfect fit (bottom center of Fig.~\ref{FIG:CAPPSIM_SINGLE}), the
narrower width of the normalized power excess distribution must be
induced from the background parametrization and subtraction
thereafter~\cite{ADMX, HAYSTAC, simple_ACTION}.

Our validations shown in Figs.~\ref{FIG:CAPPSIM_FLAT} and
\ref{FIG:CAPPSIM_SINGLE} demonstrate not only our solid understanding
of the analysis procedure but also the validity of our simulation data.

\section{Incorporating the full correlations}\label{FULLCORR}
Equation~(\ref{EQ:COADD_POWER}) shows the power excess in one of the
frequency bins of the grand power spectrum with the signal
weighting~\cite{HAYSTAC}
\begin{equation}
  P_{\varA{g}}=\sum_{i=1}^{\substack{\varA{co-adding}\\ \varA{bins}}}P_i \mathcal{L}_i
  \label{EQ:COADD_POWER}
\end{equation}
and Eq.~(\ref{EQ:COADD_CORR}) shows the full error propagation of the
power excess shown in Eq.~(\ref{EQ:COADD_POWER})
\begin{equation}
  \sigma^2_{P_{\varA{g}}} = 
  \sum_{i=1}^{\substack{\varA{co-adding}\\ \varA{bins}}}\sum_{j=1}^{\substack{\varA{co-adding}\\ \varA{bins}}}\sigma_{P_i}\sigma_{P_j}\mathcal{L}_i\mathcal{L}_j\rho_{ij},
  \label{EQ:COADD_CORR}
\end{equation}
where $P_i$ and $\sigma_{P_i}$ are the power excess and associated error
in the $i_{\rm th}$ frequency bin of the combined power spectrum after
the Step-2 procedure, respectively, and $\mathcal{L}_i$ is an axion signal
likelihood for both the $P_i$ and $\sigma_{P_i}$ weightings. With the
signal window $b_a$ of 5000 Hz as shown in Fig.~\ref{FIG:AXION_SHAPE},
$\mathcal{L}_i$ meets the condition
$\sum_{i=1}^{\substack{\varA{co-adding}\\ \varA{bins}}}\mathcal{L}_i\simeq1$
with 10 co-adding frequency bins and an RBW of 500 Hz for the CAPP-8TB
axion haloscope search. $\rho_{ij}$ is the correlation coefficient
between the frequency bins participating in the co-adding procedure,
thus is unity for $i=j$.
The narrower width of the normalized power excess distribution from
the normalized grand power spectrum could mean that the noise
fluctuations in the grand power spectrum $\sigma_{P_\varA{g}}$ are
overestimated with the terms for $i=j$ only in
Eq.~(\ref{EQ:COADD_CORR}) or, equivalently, with
$\sum_{i=1}^{\substack{\varA{co-adding}\\ \varA{bins}}}\sigma^2_{P_i}\mathcal{L}^2_i$
only.
The correlation coefficients $\rho_{ij}$ in Eq.~(\ref{EQ:COADD_CORR})
are very likely to be negative due to overestimation of the noise
fluctuations in the grand power
spectrum~\cite{ADMX, HAYSTAC, simple_ACTION}.
\begin{figure}[h]
  \centering
  \includegraphics[width=1.0\textwidth]{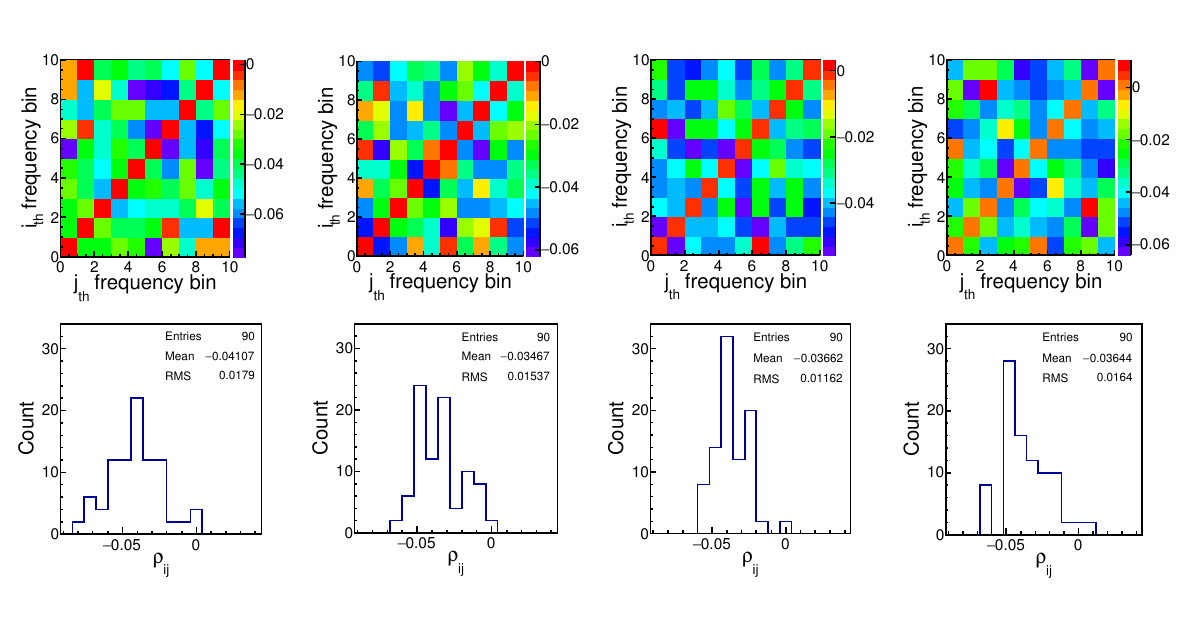}
  \caption{Examples of the correlation coefficient maps (top) and the
    projections onto the correlation coefficient axis (bottom). In the
    top 2-D plots, $\rho_{ij}$ for $i=j$ are unity with 100\% correlation,
    but are set to zero to distinguish the differences among $\rho_{ij}$
    for $i\ne j$. $\rho_{ij}$ for $i=j$  are also left out in the bottom
    histograms and that is the reason the entries in them are not 100,
    but 90 from a 10$\times$10 matrix. Plots in the 1st column were
    obtained from a frequency that has axion signals, and those in the
    next three columns from other frequencies with no axion signals.}  
  \label{FIG:CORRMATRIX}
\end{figure}

With 10 co-adding bins, each frequency bin in the
grand power spectrum has a full error contribution from $10\times10$
combinations or, equivalently, from a relevant $10\times10$
correlation matrix. The total number of frequencies in our grand power
spectrum is 100001 with a frequency band of 50 MHz (1600 to 1650 MHz)
and an RBW of 500 Hz~\cite{CAPP-8TB-PRL}, hence we need to construct
100001 $10\times10$ correlation matrices. These 100001 correlation
matrices were constructed from the CAPP-8TB simulation data and
background only simulation data. The elements in each correlation
matrix were calculated as the standard Pearson correlation
coefficient. The CAPP-8TB simulation data was used for the background
parametrizations and the parametrizations were then applied to the
background only simulation data to extract the power excess for the
correlation coefficient calculations.
Figure~\ref{FIG:CORRMATRIX} shows examples of the correlations
obtained from the simulation data. The 1st column shows the
correlation coefficient map and the distribution of the correlation
coefficients from a frequency that has input axion signals, and the
other columns show them from frequencies that have no axion
signals. As predicted earlier, most of the correlation coefficients
were constructed as negative values, which can explain the narrower
widths of the normalized power excess distributions shown in the
bottom left (experimental data) and right (simulation data) of
Fig.~\ref{FIG:CAPPSIM_SINGLE}.
\begin{figure}[h]
  \centering
  \includegraphics[width=0.5\textwidth]{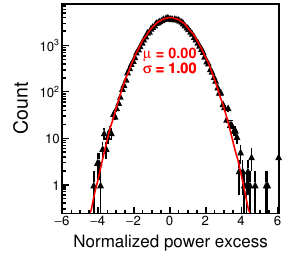}
  \caption{The figure shows the same data as in the bottom left of
    Fig.~\ref{FIG:CAPPSIM_SINGLE} but with the correlation matrices
    fully incorporated.}  
  \label{FIG:CAPP-8TB-results-data}
\end{figure}
\begin{figure}[b]
  \centering
  \includegraphics[width=1.0\textwidth]{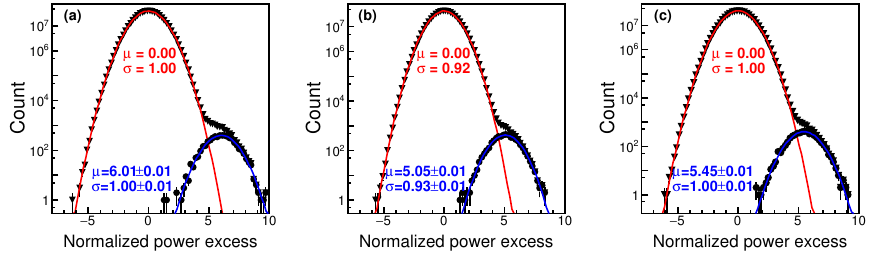}
  \caption{The inverted triangles in (a), (b), and (c) are the
    distributions of the normalized power excess from all the
    frequency bins in the normalized grand power spectra from the 5000
    simulated CAPP-8TB experiments, while the circles are those
    from a particular frequency bin where we put simulated axion
    signals on top of the CAPP-8TB background. (a) was obtained after
    subtracting the background with perfect fit; (b) and (c) with a
    5-parameter fit; (c) incorporates the correlations between the
    co-adding frequency bins. Lines are a Gaussian fit.}
  \label{FIG:CAPP-8TB-results-simulation}
\end{figure}
The 100001 correlation matrices constructed from the simulation data
were then fed through the CAPP-8TB experimental data and CAPP-8TB
simulation data in the Step-3 procedure.
Figure~\ref{FIG:CAPP-8TB-results-data} was obtained from the CAPP-8TB
experimental data~\cite{CAPP-8TB-PRL}, which corresponds to the bottom
left data in Fig.~\ref{FIG:CAPPSIM_SINGLE}. After the full
incorporation of the correlation matrices the triangles now follow a
standard Gaussian distribution.

The results from the CAPP-8TB simulation data are also shown in (a),
(b), and (c) in Fig.~\ref{FIG:CAPP-8TB-results-simulation},
respectively~\cite{CAPP-8TB-PRL}.
Figures~\ref{FIG:CAPP-8TB-results-simulation}(a) and
\ref{FIG:CAPP-8TB-results-simulation}(b) correspond to the bottom center
and right of Fig.~\ref{FIG:CAPPSIM_SINGLE}, respectively, with 5000
simulated CAPP-8TB experiments.
Figure~\ref{FIG:CAPP-8TB-results-simulation}(c) was also from 5000
simulated CAPP-8TB experiments. Both the inverted triangles and
circles follow a Gaussian distribution with a width exhibiting unity,
thanks to the full incorporation of the correlation matrices. From the
mean values of the circle distributions in
Figs.~\ref{FIG:CAPP-8TB-results-simulation}(a),
\ref{FIG:CAPP-8TB-results-simulation}(b), and
\ref{FIG:CAPP-8TB-results-simulation}(c), the SNR efficiencies with a
5-parameter fit are 90.7\% and 84.0\% with and without the correlation
matrices, respectively. The $\epsilon_{\rm SNR}$ of 90.7\% with the
full frequency-dependent correlations shows good agreement with what
the HAYSTAC achieved with a single frequency-independent numerical
factor~\cite{HAYSTAC} and was applied to our previous
publication~\cite{CAPP-8TB-PRL}. One also can expect such an
$\epsilon_{\rm SNR}$ agreement between the HAYSTAC and CAPP-8TB
experiments based on Fig.~\ref{FIG:CORRMATRIX}, where each correlation
coefficient map does not depend significantly on the frequency, or is
almost unaffected by the presence of the axion signals.

\section{Improvement}
Following the HAYSTAC method, the normalized power excess in the
Step-3 can be expressed as
\begin{equation}
  \varA{Normalized~grand~power~excess}=\frac{P_\varA{grand}}{\sigma_{P_\varA{grand}}\xi_{\varA{Step-3}}},
  \label{snrSF}
\end{equation}
where $P_\varA{grand}=P_\varA{g}$,
$\sigma^2_{P_{\varA{grand}}}=\sum_{i=1}^{\substack{\varA{co-adding}\\\varA{bins}}}\sigma^2_{P_i}\mathcal{L}^2_i$,
and $\xi_\varA{Step-3}$ is a frequency-independent scale factor to
remedy the bias induced from the background parametrizations. Its
value for the CAPP-8TB experiment is 0.92, as shown in
Fig.~\ref{FIG:CAPP-8TB-results-simulation}(b).
Another frequency-independent scale factor $\xi_\varA{Step-1.5}$ of
0.98 was used in Step-1.5 for the distributions shown in the middle left
(experimental data) and right (simulation data) of
Fig.~\ref{FIG:CAPPSIM_SINGLE}, hence for the CAPP-8TB
results~\cite{CAPP-8TB-PRL}, which is also induced by the background
parametrizations. The result using a $\xi_\varA{Step-3}$ of 0.92 is in
good agreement with that using the full correlations. Having said
that, through the rest of this work we will employ
frequency-independent scale factors.

Equation~(\ref{snrSF}) itself is valid whether the scale factor
affects $P_\varA{grand}$ or $\sigma_\varA{grand}$, though it was
applied to the latter. First, we looked into which are affected by a
5-parameter fit using the CAPP-8TB simulation
data. Table~\ref{TAB:FITS} shows the category of the normalized grand
power excess with different $P_\varA{grand}$ and $\sigma_\varA{grand}$
depending mainly on the background parametrizations.
\begin{table}[h]
  \centering
  \begin{tabular}{cll}\hline\hline
    &$P_\varA{grand}$& $\sigma_{P_\varA{grand}}$ \\ \hline
    (A) &perfect fit       & perfect fit \\ 
    (B) &perfect fit       & 5-parameter fit\\ 
    (C) &5-parameter fit& perfect fit\\ 
    (D) &5-parameter fit& 5-parameter fit\\ 
    (E) &5-parameter fit with $\xi_\varA{Step-1.5}$& 5-parameter fit\\ 
    (F) &5-parameter fit with $\xi_\varA{Step-1.5}$ and $\xi_\varA{Step-3}$ & 5-parameter fit\\ \hline\hline
  \end{tabular}
  \caption{(A) to (F) is the category of the normalized grand power
    excess with different $P_\varA{grand}$ and $\sigma_\varA{grand}$
    depending mainly on the background parametrizations.}      
  \label{TAB:FITS}
\end{table}
\begin{figure}[h]
  \centering
  \includegraphics[width=1.0\textwidth]{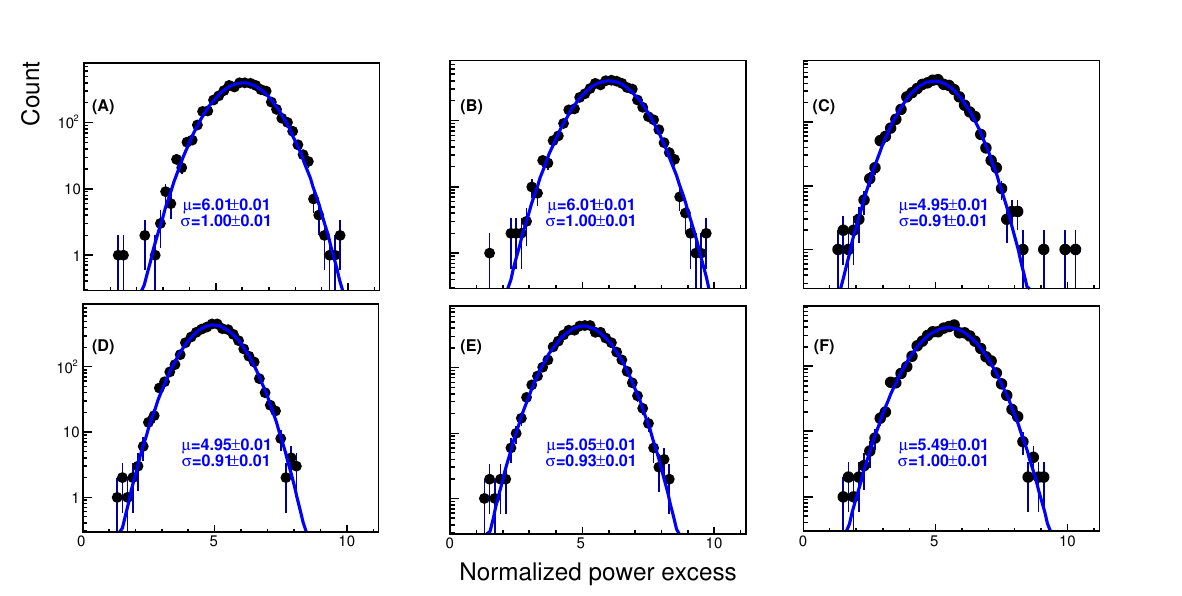}
  \caption{Normalized grand power excess distributions from the 5000
    simulated CAPP-8TB experiments at the simulated axion signal
    frequency, where the conditions for (A) to (F) are listed in
    Table~\ref{TAB:FITS}. Lines are a Gaussian fit resulting in $\mu$
    and $\sigma$.}
  \label{FIG:PCORRorECorr}  
\end{figure}
Figure~\ref{FIG:PCORRorECorr} shows the normalized grand power excess
distributions at the simulated axion signal frequency, where they are
categorized according to Table~\ref{TAB:FITS}. Since we found the
5-parameter fit actually affects only $P_\varA{grand}$ from the
simulation study shown in Table~\ref{TAB:FITS} and
Fig.~\ref{FIG:PCORRorECorr}, we will correct for $P_\varA{grand}$ only
to improve $\epsilon_{\rm SNR}$.

\begin{figure}[b]
  \centering
  \includegraphics[width=0.5\textwidth]{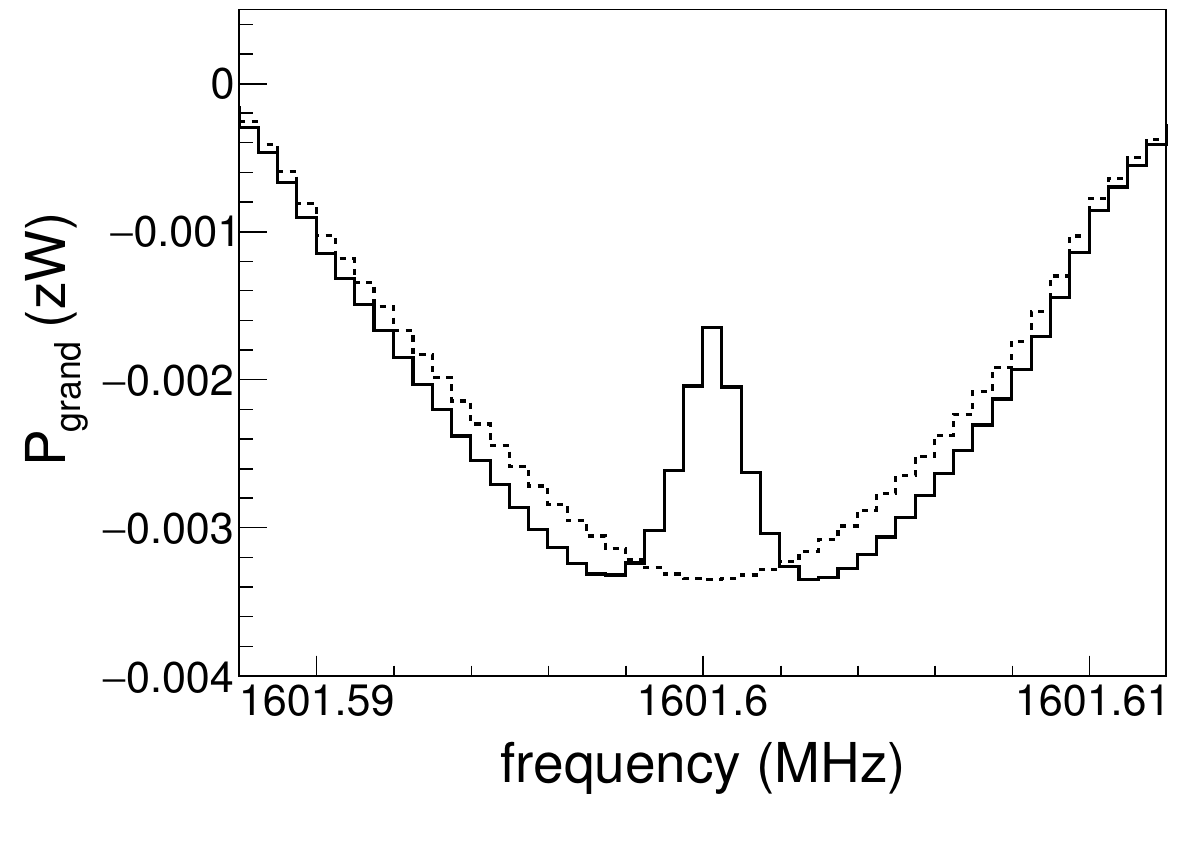}
  \caption{Differences between $P^\varA{5-parameter~fit}_\varA{grand}$
    and $P^\varA{perfect~fit}_\varA{grand}$, where the former was
    obtained with 5-parameter fit and the latter with perfect fit. The
    dashed line was obtained without $\xi_\varA{Step-1.5}$ and
    $\xi_\varA{Step-3}$, while the solid line reflects the two
    frequency-independent scale factors.}  
  \label{FIG:FITBIAS}  
\end{figure}

Second, we scrutinized the signal region and found that the
5-parameter fit little affects the axion signal power, but mainly
distorts the background in the presence of the axion
signals. The dashed lines in Fig.~\ref{FIG:FITBIAS} are the
$P_\varA{grand}$ difference between (C) and (A) in
Table~\ref{TAB:FITS} and show no axion signal excess, which implies
the axion signal power identified using the 5-parameter fit and perfect fit
are the same. The solid line in Fig.~\ref{FIG:FITBIAS} is the
$P_\varA{grand}$ difference between (F) and (A) in
Table~\ref{TAB:FITS} and the axion signal excess is not due to the
5-parameter fit, but due to the two frequency-independent scale
factors $\xi_\varA{Step-1.5}$ and $\xi_\varA{Step-3}$, which can
result in overestimated SNRs. The difference between the dashed and
solid lines in the axion signal sideband regions are also due to
$\xi_\varA{Step-1.5}$ and $\xi_\varA{Step-3}$.
\begin{figure}[h]
  \centering
  \includegraphics[width=0.7\textwidth]{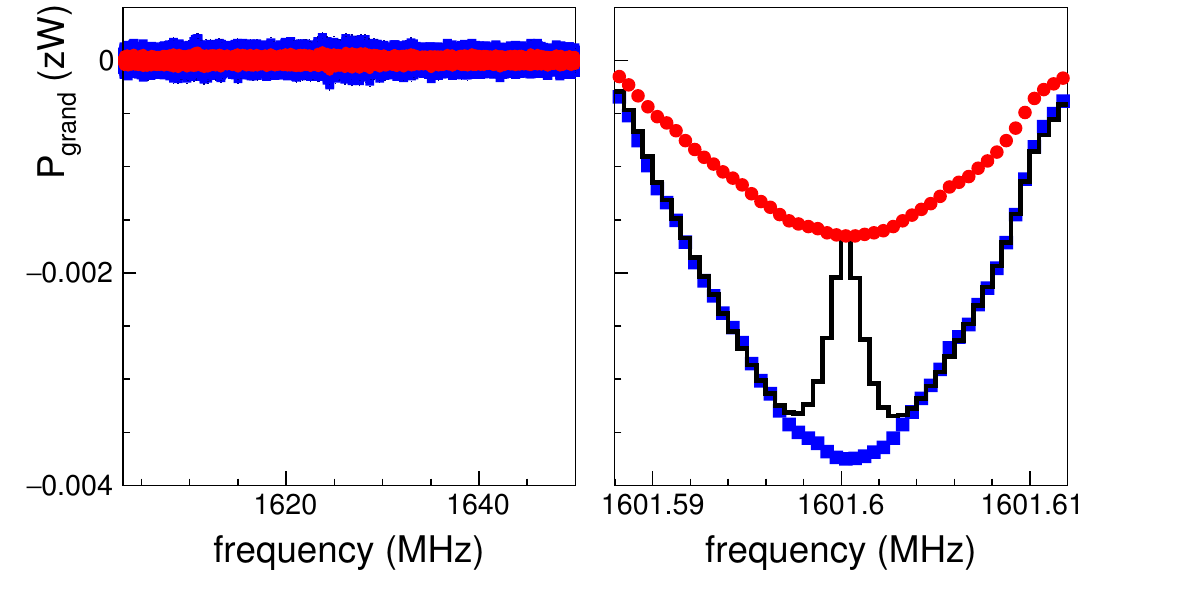}
  \caption{Rectangles (blue) and circles (red) are the underestimated
    and scaled background corrections, respectively, obtained from the
    CAPP-8TB simulation data and background only simulation data. Left
    and right show the corrections in the frequency regions without
    and with the simulated axion signals, respectively, and the solid
    line (black) on the right is the same as the solid line in
    Fig.~\ref{FIG:FITBIAS}.}  
  \label{FIG:PCORR}  
\end{figure}
\begin{figure}
  \centering
  \includegraphics[width=0.7\textwidth]{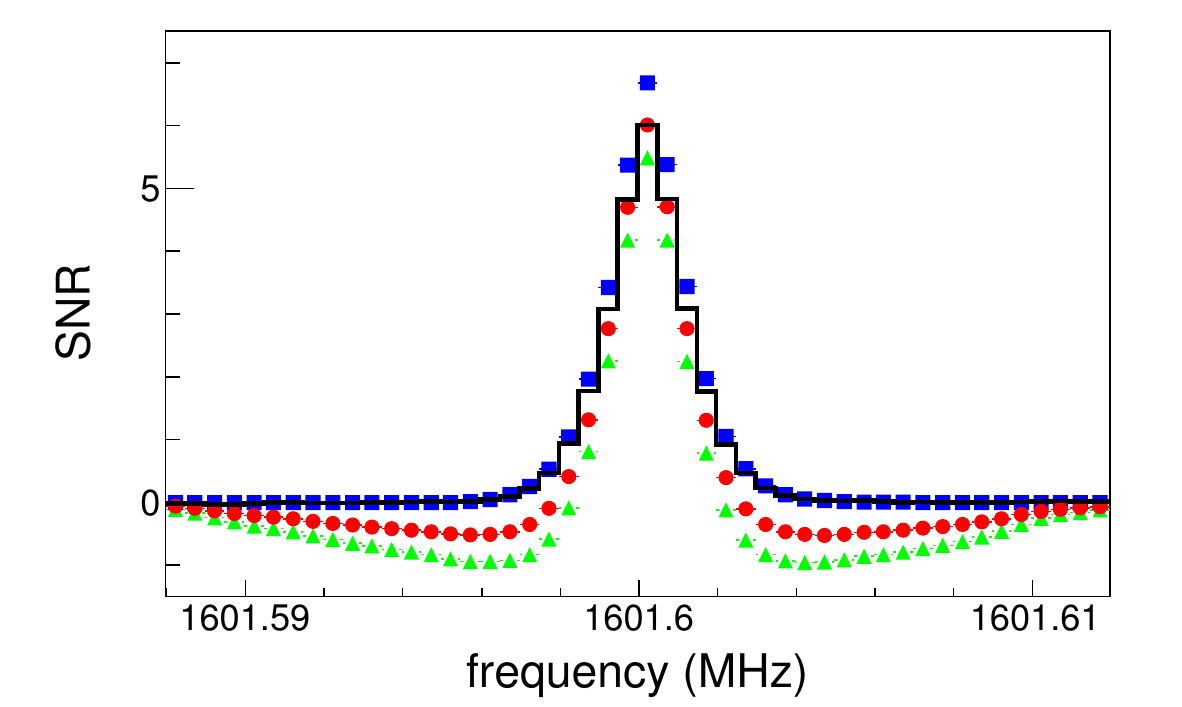}
  \includegraphics[width=0.7\textwidth]{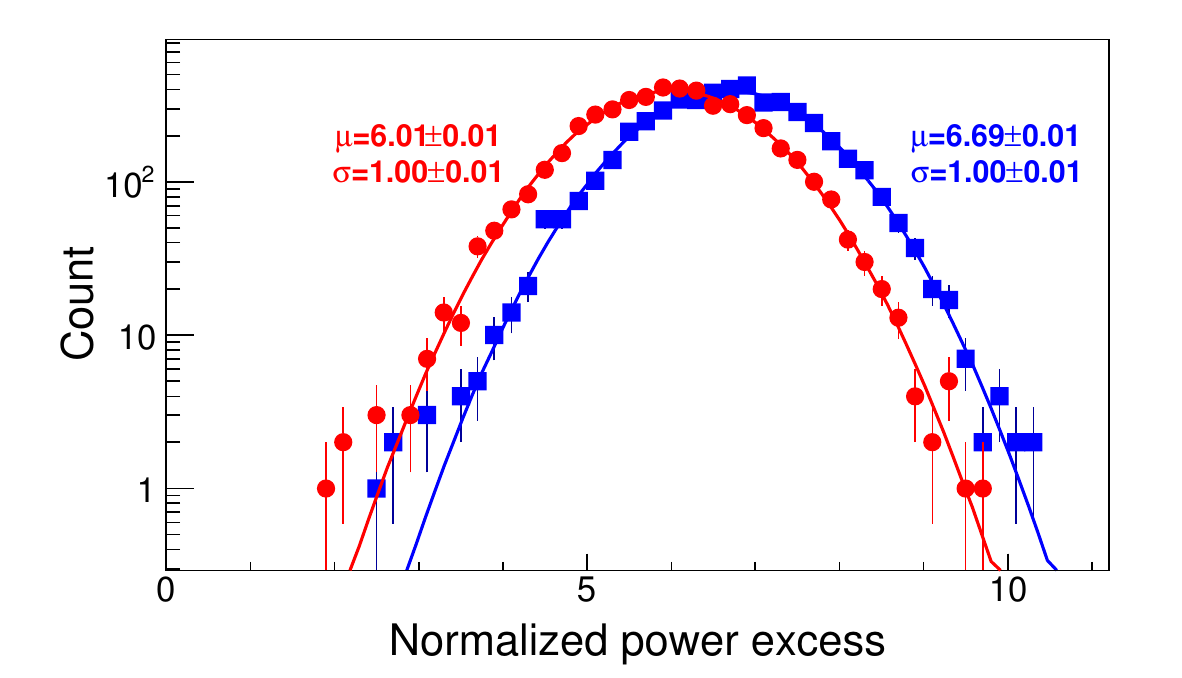}
  \caption{Top shows the final SNRs after the Step-3 procedure, as
    a function of frequency around the signal region only, from 5000
    simulated CAPP-8TB experiments. The line (black) was obtained with
    perfect fit, triangles (green) with 5-parameter fit without any
    background corrections, rectangles (blue) with 5-parameter fit
    with the underestimated background correction, and circles (red)
    with 5-parameter fit with the scaled background correction,
    respectively. The circles and rectangles in the bottom plot are
    the projections of the circle and rectangle of the top plot at the
    signal frequency which shows the best SNR. Other projections
    of the line and triangle of the top plot in the signal frequency
    correspond to (A) and (F) in Fig.~\ref{FIG:PCORRorECorr},
    respectively. Lines in the bottom plot are a Gaussian fit
    resulting in $\mu$ and $\sigma$.}  
  \label{FIG:SNR-PCORR}  
\end{figure}

We corrected for the background biased by the background
parametrizations in the presence of axion signals using the CAPP-8TB
simulation data and background only simulation data.
The CAPP-8TB simulation data was used for the background
parametrizations. The parametrizations were then applied to the
background only simulation data to extract the background corrections
shown as rectangles (blue) in Fig.~\ref{FIG:PCORR}, which are actually
the power excess for the correlation coefficient calculations in
Sec.~\ref{FULLCORR}, after further going through the Step-3
procedure. However, the background correction using the rectangles in
Fig.~\ref{FIG:PCORR} cannot suppress the overestimated axion signal
excess, which is shown in Figs.~\ref{FIG:FITBIAS} and \ref{FIG:PCORR}
by the solid line. This can result in overestimated SNRs as mentioned
earlier. Hence we refer to the correction as ``underestimated
background correction''.
To avoid the overestimated axion signal excess, we applied
another scale factor $\zeta$ of 0.44 to the underestimated background
correction, which is shown in Fig.~\ref{FIG:PCORR} as circles (red)
which are referred to as ``scaled background correction''. The scale
factor of 0.44 was obtained at the axion signal frequency by equating
the difference between the two background corrections (rectangles and
circles in Fig.~\ref{FIG:PCORR}) and the overestimated axion signal
excess (solid line in Fig.~\ref{FIG:FITBIAS}), where the scaled
background correction is the product of the scale factor and the
underestimated background correction.
The left and right plots in Fig.~\ref{FIG:PCORR} show the background
corrections in the frequency regions without and with the simulated
axion signals, respectively. The narrower band with circles
(red) in the left plot resulted from the scale factor $\zeta$. Note
that the noise power with the co-adding and signal weighting is about
6.5 zW in the signal region, thus the scaled background correction is
at most 0.03\% of the noise power. The underestimated background
correction is at most 0.06\% of the noise power, which can explain why
the $\sigma_\varA{grand}$ from the 5-parameter fit has no problems, as
shown in Fig.~\ref{FIG:PCORRorECorr}(B).
Figure~\ref{FIG:SNR-PCORR} shows the SNRs after applying the two
background corrections, where rectangles (blue) are SNRs with the
underestimated background correction and circles (red) are those with
the scaled background correction. The SNR using the scaled background
correction shows good agreement with that using perfect fit (solid
line) in the axion signal frequency, while the SNR with the
underestimated background correction is overestimated by a factor of
$1/(\xi_\varA{Step-1.5}\xi_\varA{Step-3})$, as shown in the bottom
plot of Fig.~\ref{FIG:SNR-PCORR}, as expected earlier.
Finally, we confirmed that the scale factor $\zeta$ is a
frequency-independent numerical factor from the other CAPP-8TB
simulation data, which has to be. Otherwise, the direction of the
background correction in this paper is undesirable for axion haloscope
searches because of the unknown axion mass.

\section{Summary}
We report an approach to improve axion haloscope search analyses using
the data obtained with the CAPP-8TB haloscope. By correcting for the
background biased by the background parametrizations in the presence
of axion signals, we realized an $\epsilon_\varA{SNR}$ of about
100\%. Given the axion haloscope search analyses to date, the scanning
rate can be improved by 21\%, with about a 10\% improvement in the
SNR.
For the CAPP-8TB results~\cite{CAPP-8TB-PRL}, the limits of
$g_{a\gamma\gamma}$ can be improved by about 5\% with an
improved $\epsilon_{\rm SNR}$, e.g., from about 1.00$\times 10^{-14}$
to 0.95$\times 10^{-14}$ GeV$^{-1}$.
This improvement is another low cost innovation in axion haloscope
searches, where all the experimental parameters are currently at their
best.

\acknowledgments
This work is supported by the Institute for Basic Science (IBS) under
Project Code No. IBS-R017-D1-2021-a00.


\begin{thebibliography}{99}
\bibitem{SMALLERROR}
  Throughout this paper, negligible fitting errors of $\mathcal{O}(10^{-3})$ were left out.
\end{thebibliography}
\end{document}